\documentclass[]{svjour2}

\smartqed

\usepackage{amssymb,amsmath}
\usepackage{graphicx}

\journalname{J Stat Phys}

\begin{document}

\title{Fourth moment of the charge density induced around a guest charge
in two-dimensional jellium}

\titlerunning{Fourth moment of the charge density around a guest charge}

\author{Ladislav \v{S}amaj}

\institute{Institute of Physics, Slovak Academy of Sciences, 
D\'ubravsk\'a cesta 9, SK-84511 Bratislava, Slovakia \\
\email{Ladislav.Samaj@savba.sk}}

\date{Received:  / Accepted: }

\maketitle

\begin{abstract}
The model under consideration is the classical two-dimensional one-component 
plasma (jellium) of pointlike particles with charge $e$, interacting pairwisely
via the logarithmic Coulomb potential and immersed in a uniform neutralizing 
background charge density.
The system is in thermal equilibrium at the inverse temperature $\beta$,
its thermodynamics depends only on the coupling constant $\Gamma=\beta e^2$.
We put into an infinite (homogeneous and translationally invariant)
plasma a guest particle of charge $Ze$ with $Z>-2/\Gamma$ in order to prevent
from the collapse of the jellium charges onto it. 
The guest particle induces a screening cloud (the excess charge density)
in the plasma.
The zeroth and second moments of this screening cloud were derived previously
for any fluid value of $\Gamma$.
In this paper, we propose a formula for the fourth moment of 
the screening cloud.
The derivation is based on the assumption that the fourth moment is,
similarly as the second moment, analytic in $Z$ around $Z=0$. 
An exact treatment of the limit $Z\to\infty$ shows that it is a finite (cube) 
polynomial in $Z$. 
The $\Gamma$-dependence of the polynomial coefficients is determined uniquely 
by considering the limits $Z\to 0$ and $Z\to\infty$, and the compressibility 
sum rule for $Z=1$.
The formula for the fourth moment of screening cloud is checked in the leading 
and first correction orders of the Debye-H\"uckel limit $\Gamma\to 0$ and 
at the exactly solvable free-fermion point $\Gamma=2$.
Sufficient conditions for sign oscillations of the induced charge density 
which follow from the second-moment and fourth-moment sum rules are discussed. 

\keywords{Coulomb fluids\and Jellium\and Logarithmic interaction\and Sum rules}

\end{abstract}

\renewcommand{\theequation}{1.\arabic{equation}}
\setcounter{equation}{0}

\section{Introduction} \label{Sect1}
Systems of classical (i.e. non-quantum) particles interacting via the Coulomb 
potential are fundamental models of equilibrium statistical mechanics.
To mimic adequately the standard $1/r$ potential in three dimensions (3D),
the Coulomb potential in spatial dimension $d$ is defined as the solution
of the $d$-dimensional Poisson equation with given boundary conditions. 
The long-range tail of the Coulomb potential implies exact constraints
(sum rules) on the particle correlations, see reviews \cite{Attard02,Martin88}.
Thermal equilibrium of Coulomb systems is exactly solvable in 
the high-temperature region within the linear Debye-H\"uckel mean-field 
theory \cite{Attard02,Baus80}.

A simplification of realistic systems of atomic nuclei and electrons is 
represented by the one-component plasma (OCP) (sometimes referred 
to as jellium), i.e. a system of equivalent pointlike particles of 
(say elementary) charge $e$ immersed in a uniform neutralizing background 
charge density.
Although the model is usually studied in three spatial dimensions, its
two-dimensional (2D) version is of practical importance in the theory
of polyelectrolytes. 
In the 2D Euclidean space of points ${\bf r}=(x,y)$, the Coulomb potential 
$\phi$ at point ${\bf r}$, induced by a unit charge at the origin ${\bf 0}$, 
is the solution of the 2D Poisson equation and reads as
\begin{equation}
\phi({\bf r}) = - \ln \left( \frac{r}{L} \right) .
\end{equation}  
Here, $r\equiv \vert {\bf r}\vert$ and $L$ is a free length scale which
will be set for simplicity to unity.
The particles can be represented as infinitely long parallel charged lines in
the 3D space which are perpendicular to the considered 2D plane.
In the 2D Fourier space, defined by
\begin{eqnarray} 
f(r) & = & \int \frac{{\rm d}^2k}{2\pi} \tilde{f}(k) 
\exp\left( {\rm i}{\bf k}\cdot{\bf r} \right) , \label{direct} \\
\tilde{f}(k) & = & \int \frac{{\rm d}^2r}{2\pi} f(r) 
\exp\left( - {\rm i}{\bf k}\cdot{\bf r} \right) , \nonumber \\
& = & \sum_{j=0}^{\infty} \frac{(-1)^j}{(j!)^2} \left( \frac{k^2}{4} \right)^j
\int \frac{{\rm d}^2r}{2\pi} r^{2j} f(r) , \label{Fourier} 
\end{eqnarray}
the Coulomb potential has the characteristic form
\begin{equation}
\tilde{\phi}(k) = \frac{1}{k^2}
\end{equation}
with singularity at $k=0$.
The interaction energy of two particles with charge $e$ at distance $r$ 
equals to $e^2 \phi(r)$.

The 2D OCP is considered to be in thermal equilibrium at the temperature $T$,
or the inverse temperature $\beta=1/(k_{\rm B}T)$.
Due to the presence of the rigid neutralizing background, the jellium system 
is studied in the canonical ensemble with the overall charge neutrality.
The thermodynamics depends only on the coupling constant $\Gamma=\beta e^2$.
Besides the Debye-H\"uckel limit $\Gamma\to 0$, the 2D jellium is exactly 
solvable at $\Gamma=2$ by mapping onto free fermions
\cite{Alastuey81,Jancovici81a}. 
The solvable cases involve the bulk regime as well as inhomogeneous, 
semi-infinite \cite{Jancovici82} or fully finite geometries, 
see reviews \cite{DiFrancesco94,Forrester98,Jancovici92}. 
For a finite number of particles in a domain, 
an exact analytic treatment of the partition function and pair correlation 
functions is possible also at couplings $\Gamma=4$ and $\Gamma=6$, 
by using expansion in a monomial base via Jack polynomials 
\cite{Salazar16,Tellez99,Tellez12} or by mapping onto a one-dimensional chain 
of interacting anticommuting variables \cite{Samaj95,Samaj04}; the relation 
between the two methods was discussed in Ref. \cite{Mora15}.
However, in contrast to the exact solution at coupling $\Gamma=2$
which can be performed in the thermodynamic limit, the treatment
of the jellium at $\Gamma=4, 6$ becomes very complicated for relatively small 
number of particles and there is no evidence for a simplification
of the formalism in the thermodynamic limit.    

Let the canonical averaging be represented as $\langle \cdots \rangle$.
The particles will be denoted by $j=1,2,\ldots$ and their position
vectors by ${\bf r}_j$.
The microscopic total number density of particles at point ${\bf r}$ is 
defined by $\hat{n}({\bf r}) = \sum_j \delta({\bf r}-{\bf r}_j)$ and the
microscopic total (i.e. particles plus background) charge density
$\hat{\rho}({\bf r}) = e \hat{n}({\bf r}) + \rho_b$.
At the one-particle level, one defines the average number density
\begin{equation}
n({\bf r}) = \langle \hat{n}({\bf r}) \rangle 
\end{equation}
and the total charge density
\begin{equation}
\rho({\bf r}) = e n({\bf r}) + \rho_b .
\end{equation}  
At the two-particle level, one introduces the two-body density
\begin{equation}
n^{(2)}({\bf r},{\bf r}') = \left\langle \sum_{j\ne k} \delta({\bf r}-{\bf r}_j)
\delta({\bf r}'-{\bf r}_k) \right\rangle =
\langle \hat{n}({\bf r}) \hat{n}({\bf r}') \rangle
- n({\bf r}) \delta({\bf r}-{\bf r}') .
\end{equation}
One considers also the (truncated) pair correlation function
\begin{equation}
h({\bf r},{\bf r}') = \frac{n^{(2)}({\bf r},{\bf r}')}{n({\bf r})n({\bf r}')}
-1  
\end{equation}
which goes to 0 when $\vert {\bf r}-{\bf r}'\vert\to\infty$.
As concerns the correlation of the microscopic charge densities, one defines 
the structure function
\begin{eqnarray}
S({\bf r},{\bf r}') & = & 
\langle \hat{\rho}({\bf r}) \hat{\rho}({\bf r}') \rangle
- \langle \hat{\rho}({\bf r}) \rangle \langle \hat{\rho}({\bf r}') \rangle
\nonumber \\ & = & e^2 \left[ n({\bf r}) n({\bf r}') h({\bf r},{\bf r}') 
+ n({\bf r}) \delta({\bf r}-{\bf r}') \right] . \label{Sr}
\end{eqnarray}

For an infinite Euclidean surface, the mean density of particles is
constant, $n({\bf r}) = n$ and $\rho_b=-e n$, so the electroneutrality
$\rho({\bf r})=0$ is local. 
The particle density $n$ is the parameter which scales appropriately
the distance.
The two-body quantities are translational invariant, i.e. depend on
the distance between the two points: 
$n^{(2)}({\bf r},{\bf r}') = n^{(2)}(\vert {\bf r}-{\bf r}'\vert)$,
$h({\bf r},{\bf r}') = h(\vert {\bf r}-{\bf r}'\vert)$,
$S({\bf r},{\bf r}') = S(\vert {\bf r}-{\bf r}'\vert)$.   
The thermodynamics of the infinite system depends only on 
the coupling constant $\Gamma=\beta e^2$.

The long-range character of the Coulomb potential implies
exact constraints for the moments of the truncated two-body correlations.
The standard sum rules, having their counterparts in any dimension, 
are the zeroth-moment (perfect screening) condition
\begin{equation} \label{sr1}
n \int {\rm d}^2r h(r) = -1 ,
\end{equation}
the second-moment (Stillinger-Lovett) condition 
\cite{Stillinger68a,Stillinger68b}
\begin{equation} \label{sr2}
n \left( \frac{\pi\Gamma n}{2} \right) \int {\rm d}^2r r^2 h(r) = -1 
\end{equation}
and the fourth-moment (compressibility) condition 
\cite{Baus78,Vieillefosse75,Vieillefosse85}
\begin{equation} \label{sr3}
n \left( \frac{\pi\Gamma n}{2} \right)^2 \int {\rm d}^2r r^4 h(r) = 
\Gamma - 4 
\end{equation}
which is available explicitly due to the knowledge of the equation of
state for the pressure.
The sixth-moment sum rule was derived in Ref. \cite{Kalinay00} based
on ``cancellation properties'' of certain families of $c$-diagrams
generated from a renormalized Mayer diagrammatic expansion in density,
\begin{equation} \label{sr4}
n \left( \frac{\pi\Gamma n}{2} \right)^3 \int {\rm d}^2r r^6 h(r) = 
\frac{3}{4} (\Gamma-6) (8-3\Gamma) . 
\end{equation}
With regard to the definition of the Fourier transform (\ref{Fourier}),
these sum rules can be represented as the small-$k$ expansion of the
Fourier transform of the correlation function  
\begin{eqnarray}
2\pi n \tilde{h}(k) & = & -1 + \frac{k^2}{2\pi\Gamma n}
+ \left( \frac{\Gamma}{4} - 1 \right) \left( \frac{k^2}{2\pi\Gamma n} \right)^2
\nonumber \\ & &
+ \left( \frac{\Gamma}{4}-\frac{3}{2} \right)
\left( \frac{\Gamma}{4}-\frac{2}{3} \right)
\left( \frac{k^2}{2\pi\Gamma n} \right)^3 + O(k^8) . \label{defh}
\end{eqnarray}
According to the definition (\ref{Sr}) of the structure function,
its Fourier transform is related to $\tilde{h}(k)$ as follows.
\begin{equation} \label{defS}
\frac{\tilde{S}(k)}{e^2} = \frac{n}{2\pi} + n^2 \tilde{h}(k) . 
\end{equation}

The homogeneity and translational symmetry of the infinite 2D OCP can 
be broken by putting a ``guest'' pointlike particle with charge $Z e$ into the
bulk interior of the plasma, say at the origin ${\bf 0}$ of the coordinate 
system. 
The Boltzmann factor of the guest charge $Z e$ with a jellium charge
$e$ at distance $r$, $r^{\Gamma Z}$, can be integrated at small distances $r$
in 2D only if
\begin{equation} \label{con}
Z > - \frac{2}{\Gamma} .
\end{equation} 
This lower bound for $Z$ prevents from the thermodynamic collapse of 
the jellium charge onto the guest charge.
Let $n({\bf r}\vert Ze,{\bf 0})$ denotes the averaged density of jellium 
charges in the presence of the guest charge $Ze$ at the origin ${\bf 0}$.
The corresponding (total) charge density is
$\rho({\bf r}\vert Ze,{\bf 0}) = e [n({\bf r}\vert Ze,{\bf 0}) - n]$.
If $Z=0$ we have trivially
\begin{equation} \label{Z0}
\rho({\bf r}\vert 0,{\bf 0}) = 0 . 
\end{equation}
If $Z=1$, the fixed particle is the one forming the jellium and
we have the obvious identification
\begin{equation}
n({\bf r}\vert e,{\bf 0}) = \frac{n^{(2)}({\bf 0},{\bf r})}{n({\bf 0})} ,
\end{equation}
where $n({\bf 0}) = n$.
The plasma sum rules (\ref{sr1})--(\ref{sr4}) then take the form 
\begin{eqnarray} 
\int {\rm d}^2r \rho({\bf r}\vert e,{\bf 0})  & = & - e , \label{srr1} \\
\int {\rm d}^2r r^2 \rho({\bf r}\vert e,{\bf 0}) & = & -e 
\left( \frac{2}{\pi\Gamma n} \right) , \label{srr2} \\
\int {\rm d}^2r r^4 \rho({\bf r}\vert e,{\bf 0}) & = & e 
\left( \frac{2}{\pi\Gamma n} \right)^2 (\Gamma - 4) , \label{srr3} \\
\int {\rm d}^2r r^6 \rho({\bf r}\vert e,{\bf 0})  & = &
e \left( \frac{2}{\pi\Gamma n} \right)^3 \frac{3}{4} 
(\Gamma-6) (8-3\Gamma) . \label{srr4}
\end{eqnarray}  

As concerns the known results for any value of $Z$ constrained by (\ref{con}),
the generalization of the zeroth-moment relation (\ref{srr1}) is obvious,
\begin{equation} \label{srrr1}
\int {\rm d}^2r \rho({\bf r}\vert Ze,{\bf 0})  = - Z e , 
\end{equation}
i.e. the charge cloud around the guest charge carries exactly the
opposite charge $-Z e$ to maintain the overall electroneutrality.
The generalization of the second-moment relation (\ref{srr2}) 
\begin{equation} \label{srrr2} 
\int {\rm d}^2r r^2 \rho({\bf r}\vert Ze,{\bf 0}) = -Z e 
\left( \frac{2}{\pi\Gamma n} \right) 
\left[ \left( 1 - \frac{\Gamma}{4} \right) + \frac{\Gamma}{4} Z \right]
\end{equation}
was made in Ref. \cite{Samaj07} by using a mapping technique of the 2D OCP
onto a discrete one-dimensional anticommuting-field theory.
The rederivation of this result and its generalization to a 2D mixture 
of several species by using the BGY hierarchy was done in Ref. 
\cite{Jancovici08}.
The extension of the analysis to ionic mixtures in higher spatial dimensions 
was the subject of Ref. \cite{Suttorp08}.

The aim of this paper is to derive a formula for the fourth moment of 
the screening cloud around the guest particle with charge $Z e$.
We proceed in close analogy with Sect. 2.3 of Ref. \cite{Jancovici08}
dealing with the second moment of the charge cloud where an assumption 
was made that this second moment can be expanded in integer powers of $Z$,
i.e. it is analytic in $Z$ around $Z=0$. 
We apply this assumption also to the fourth moment and derive the expansion 
coefficients by using exact treatment of the limits $Z\to 0$ and $Z\to\infty$,
and the known compressibility sum rule for $Z=1$.
The final result reads as
\begin{subequations} \label{fourth}
\begin{equation} \label{fourtha}
\frac{(\pi\Gamma n)^2}{e} \int {\rm d}^2r r^4 \rho({\bf r}\vert Ze,{\bf 0}) 
= c_1 Z + c_2 Z^2 + c_3 Z^3 ,
\end{equation}
where
\begin{eqnarray} 
c_1(\Gamma) & = & - \left( \Gamma-6 \right) 
\left( \Gamma-\frac{8}{3} \right) , \nonumber \\
c_2(\Gamma) & = & \frac{2}{3} \Gamma (2\Gamma-7) , \nonumber \\
c_3(\Gamma) & = & - \frac{1}{3} \Gamma^2 . \label{fourthb} 
\end{eqnarray}
\end{subequations}
The absolute term of order $Z^0$ is equal to 0 due to the equality
(\ref{Z0}).
The formula is checked in the leading and first correction orders of
the Debye-H\"uckel limit $\Gamma\to 0$ and at the exactly solvable 
free-fermion point $\Gamma=2$.

The derivation of the fourth-moment condition (\ref{fourth}) is presented 
in Sect. \ref{Sect2}.
It is based on an exact treatment of the limits $Z\to 0$ (Sect. \ref{Sect21}) 
and $Z\to\infty$ (Sect. \ref{Sect22}), and the known compressibility sum rule 
for $Z=1$ (Sect. \ref{Sect23}).
The formula is checked in the leading and first correction orders of
the Debye-H\"uckel limit $\Gamma\to 0$ in Sect. \ref{Sect3}.
The check of the formula at the free-fermion point $\Gamma=2$ 
is given in Sect. \ref{Sect4}.
Sufficient conditions for sign oscillations of the induced charge density 
which follow from the second-moment and fourth-moment sum rules are discussed 
in Sect. \ref{Sect5}.
Sect. \ref{Sect6} brings concluding remarks.

\renewcommand{\theequation}{2.\arabic{equation}}
\setcounter{equation}{0}

\section{Derivation of the fourth moment} \label{Sect2}

\subsection{Limit $Z\to 0$} \label{Sect21}
Introducing the guest charge $Ze$ at the origin ${\bf 0}$, one adds
to the microscopic plasma Hamiltonian the additional part
\begin{equation}
\hat{H}' = Z e \hat{\phi}({\bf 0}) ,
\end{equation}
where
\begin{equation}
\hat{\phi}({\bf 0}) = \int {\rm d}^2r' \phi({\bf r}') \hat{\rho}({\bf r}')
\end{equation}
is the microscopic electric potential induced by the OCP at the origin.
The excess charge density at point ${\bf r}$ is given by
\begin{equation} \label{rho}
\rho({\bf r}\vert Ze,{\bf 0}) = \frac{\langle \hat{\rho}({\bf r}) 
\exp(-\beta \hat{H}') \rangle}{\langle \exp(-\beta \hat{H}') \rangle} ,
\end{equation}
where $\langle \cdots \rangle$ denotes the canonical averaging over
the infinite (homogeneous and isotropic) plasma system.
Within linear response theory, expanding the exponentials in (\ref{rho}) 
to first order in $-\beta \hat{H}'$ the charge density (\ref{rho}) can be 
expanded to first order in $Z$ as follows   
\begin{equation}
\rho({\bf r}\vert Ze,{\bf 0}) \mathop{\sim}_{Z\to 0}
- Z e \beta \int {\rm d}^2 r' S({\bf r},{\bf r}') \phi({\bf r}') . 
\end{equation}
Within the Fourier representation (\ref{direct}) and (\ref{Fourier}),
by using the convolution theorem we obtain that
\begin{equation} \label{linZ}
\tilde{\rho}({\bf k}\vert Ze,{\bf 0}) \mathop{\sim}_{Z\to 0} - Z e (2\pi) \beta 
\tilde{\phi}(k) \tilde{S}(k) = - Z e \frac{2\pi\beta}{k^2} \tilde{S}(k) .
\end{equation}
Formula for the small-$k$ expansion of $\tilde{S}(k)$ results from 
the combination of Eqs. (\ref{defh}) and (\ref{defS}).
Consequently, 
\begin{eqnarray}
\tilde{\rho}({\bf k}\vert Ze,{\bf 0}) & \displaystyle{\mathop{\sim}_{Z\to 0}} & 
- \frac{Z e}{2\pi} \left[ 1 + \left( \frac{\Gamma}{4}-1 \right) 
\frac{k^2}{2\pi\Gamma n} \right. \nonumber \\ & & \left.
+ \left( \frac{\Gamma}{4}-\frac{3}{2} \right) 
\left( \frac{\Gamma}{4}-\frac{2}{3} \right) 
\left( \frac{k^2}{2\pi\Gamma n} \right)^2 + O(k^6) \right] . 
\end{eqnarray}
With respect to the general form of the $k^2$-expansion (\ref{Fourier}) for 
the 2D Fourier transform, we finally arrive at
\begin{equation} \label{Zlinear} 
\frac{(\pi\Gamma n)^2}{e} \int {\rm d}^2 r r^4
\rho({\bf r}\vert Ze,{\bf 0}) \mathop{\sim}_{Z\to 0}
- Z \left( \Gamma-6 \right) \left( \Gamma-\frac{8}{3} \right) .
\end{equation}
This is the term of order $Z$ in the expansion (\ref{fourth}). 

\subsection{Limit $Z\to\infty$} \label{Sect22}
The guest particle with a large charge $Z e$ expels the mobile plasma
charges $e$ from its neighborhood region which can be taken as an isotropic 
disk of large radius $R$.
Then we can assume in the first approximation that there is only 
the background charge for $r<R$, while plasma charges fully compensate 
the background charge for $r>R$, i.e.
\begin{equation} \label{aa}
\rho({\bf r}\vert Ze,{\bf 0}) \mathop{\sim}_{Z\to\infty} \left\{
\begin{array}{cll}
-n e & & \mbox{for $r<R$,} \cr 
& & \cr
0 & & \mbox{for $r>R$.}
\end{array} \right.
\end{equation}
The radius $R$ is determined by the screening condition (\ref{srrr1})
as follows
\begin{equation}
\pi R^2 n = Z .
\end{equation} 
Thus
\begin{equation}
\int {\rm d}^2 r r^4 \rho({\bf r}\vert Ze,{\bf 0}) \mathop{\sim}_{Z\to\infty}
(-ne) 2\pi \int_0^R {\rm d}r r^5 = - Z^3 \frac{e}{3 (\pi n)^2} 
\end{equation}
which corresponds to the term of order $Z^3$ in the expansion (\ref{fourth}).   
Note that this term does not depend on $\Gamma$.

The above treatment of the limit $Z\to\infty$ has to be taken with caution
because there is a transition region of width $\propto 1/\sqrt{n}$ around $r=R$ 
where the total charge density ranges between the values $-n e$ and 0 
\cite{Jancovici81b}. 
To estimate its contribution to the fourth moment, we define
\begin{equation} 
\rho({\bf r}\vert Ze,{\bf 0}) = -n e \theta(R-r) + f_R(r) , 
\end{equation}
where $\theta$ is the Heaviside step function and $f_R(r)$ reflects
the deviation of the exact total density profile from the proposed one
in equation (\ref{aa}).
The electroneutrality condition (\ref{srrr1}) corresponds to the constraint
\begin{equation} \label{aneutr}
\int_0^{\infty} {\rm d}^2 r f_R(r) = 0 .
\end{equation}
The curvature effects become negligible in the considered limit $R\to\infty$.
Let us shift the reference point at $r=0$ to the disk surface via 
the transformation $r=R+x$; since $R\to\infty$ the coordinate $x$ covers 
the whole real axis. 
We assume that
\begin{equation}
f_R(R+x) \mathop{\sim}_{R\to\infty} F(x) + o(1) ,
\end{equation}
where $F(x)$ is the transition density profile near a rectilinear wall 
depending only on the coupling constant $\Gamma$ (and not on $Z$) and the terms
in $o(1)$ vanish in the limit $R\to\infty$.
Rewriting the left-hand-side of the constraint (\ref{aneutr}) as
\begin{equation} 
2\pi \int_0^{\infty} {\rm d} r r f_R(r) = 2\pi \int_{-R}^{\infty} {\rm d}x
(R+x) \left[ F(x) + o(1) \right] ,
\end{equation}  
in the limit $R\to\infty$ one arrives at
\begin{equation}
\int_{-\infty}^{\infty} {\rm d} x F(x) = 0 .
\end{equation}   
The contribution of the function $f_R(r)$ to the fourth moment of the total 
charge density is then given by
\begin{eqnarray} 
\int_0^{\infty} {\rm d}^2 r r^4 f_R(r) & = & 2\pi \int_{-R}^{\infty} {\rm d}x
(R+x)^5 \left[ F(x) + o(1) \right] \nonumber \\ 
& \displaystyle{\mathop{\sim}_{R\to\infty}} & 10\pi R^4 
\int_{-\infty}^{\infty} {\rm d} x x F(x) + o(R^4) .
\end{eqnarray}
Since the integral depends only on $\Gamma$, the transition function 
contributes by the lower-order correction $Z^2$ which justifies 
the above large-$Z$ analysis.
Comparing with the basic formula (\ref{fourth}) we conclude that there
should hold
\begin{equation} \label{newformula}
c_2(\Gamma) = \frac{10\pi\Gamma^2}{e} \int_{-\infty}^{\infty} {\rm d} x x F(x) .
\end{equation}
It is difficult to derive microscopically $c_2(\Gamma)$ from this relation 
for our specific type of the transition region; we shall rather determine 
the coefficient $c_2(\Gamma)$ by considering the exactly solvable $Z=1$ case
in the next subsection.
In any case, the proposed formula (\ref{newformula}) represents an alternative
way of determining $c_2(\Gamma)$ and at the same time, based on the present
suggestion for $c_2(\Gamma)$, a sum rule for the first moment of $F(x)$ for 
any coupling $\Gamma$.
 
\subsection{$Z=1$} \label{Sect23}
For $Z=1$, the moment relation (\ref{srr3}) implies that
\begin{equation} \label{Zequal}
c_1(\Gamma) + c_2(\Gamma) + c_3(\Gamma) = 4(\Gamma-4)  
\end{equation} 
which determines the coefficient of the term of order $Z^2$ in the expansion 
(\ref{fourth}).   

\renewcommand{\theequation}{3.\arabic{equation}}
\setcounter{equation}{0}

\section{Check in the Debye-H\"uckel limit} \label{Sect3}
In the Debye-H\"uckel limit $\Gamma\to 0$, the expansion of the right-hand 
side of the fourth-moment condition (\ref{fourth}) up to terms linear in 
$\Gamma$ yields
\begin{equation} \label{res}
\frac{(\pi\Gamma n)^2}{e} \int {\rm d}^2r r^4 \rho({\bf r}\vert Ze,{\bf 0}) 
= Z \left( -16 + \frac{26}{3} \Gamma \right) + Z^2 \left( -\frac{14}{3}\right)
\Gamma + O(\Gamma^2) .
\end{equation}
In this section, we derive this expansion by using a renormalized Mayer
expansion in density, initiated in Refs. \cite{Abe59,Friedman59,Meeron61}
and developed further in Refs. \cite{Deutsch74,Jancovici00b,Kalinay00,Samaj00}.

Let us consider a mixture of species $\sigma=1,2,\ldots$ with charges
$q_{\sigma}$ and homogeneous densities $n_{\sigma}$.
The background charge density $\rho_b$ is given by the condition of
overall electroneutrality as follows
\begin{equation}
\sum_{\sigma} n_{\sigma} q_{\sigma} + \rho_b = 0 .
\end{equation} 
We shall also need the definition of the inverse Debye length $\kappa$,
\begin{equation}
\kappa^2 = 2\pi\beta \sum_{\sigma} n_{\sigma} q_{\sigma}^2 .
\end{equation}
The (translational invariant) correlation functions of species $\sigma$ and 
$\sigma'$ at distance $r$ will be denoted by $h_{\sigma\sigma'}(r)$.

The Coulomb bonds $-\beta \phi(r) q_{\sigma} q_{\sigma'}$ of the standard Mayer 
expansion in density can be replaced via a series resummation by the 
renormalized bonds $-\beta K_0(\kappa r) q_{\sigma} q_{\sigma'}$ where $K_0$
is the modified Bessel function of second kind.
In contrast to the long-ranged Coulomb bonds, integrals over the renormalized 
bonds in Mayer diagrams are finite since $K_0(\kappa r)$ is short-ranged,
it goes to 0 exponentially quickly as $r\to\infty$. 
The crucial quantity of the diagrammatic method is the direct correlation 
function $c_{\sigma\sigma'}(r)\equiv c_{\sigma'\sigma}(r)$ whose expansion in 
$\beta$ can be written as
\begin{equation} \label{directcor}
c_{\sigma\sigma'}(r) = - \beta \phi(r) q_{\sigma} q_{\sigma'} + \frac{1}{2} \beta^2
K_0^2(\kappa r) q_{\sigma}^2 q_{\sigma'}^2 + O(\beta^3) .  
\end{equation}
The first term on the right-hand side of this equation corresponds to 
the original one-bond diagram which cannot be renormalized, the second one 
is the renormalized two-bond Meeron ``watermelon'' diagram.
All other renormalized diagrams possess at least three renormalized bonds
and therefore contribute to $O(\beta^3)$.
We emphasize that the classification of the renormalized diagrams in powers of 
$\beta$ (or $\Gamma$) according to their topology is a special feature 
of 2D pointlike charges with logarithmic pair interactions. 

To obtain the Fourier transform of the direct correlation function 
(\ref{directcor}), one needs the Fourier transform of $K_0^2(\kappa r)$:
\begin{eqnarray}
\int \frac{{\rm d}^2 r}{2\pi} {\rm e}^{{\rm i}{\bf k}\cdot{\bf r}} K_0^2(\kappa r)
& = & \frac{1}{\kappa^2} \int {\rm d}r' r' J_0(kr'/\kappa) K_0^2(r')
\nonumber \\ & = & \frac{1}{\kappa k} \frac{\ln\left\{ k/(2\kappa)
+ \sqrt{1+\left[k/(2\kappa)\right]^2} \right\}}{
\sqrt{1+\left[k/(2\kappa)\right]^2}} \nonumber \\ 
& \displaystyle{\mathop{\sim}_{k\to 0}} & \frac{1}{2\kappa^2}
- \frac{k^2}{12\kappa^2} + O(k^4) . 
\end{eqnarray}
Consequently, the Fourier transform of the direct correlation function
can be expressed as the small-$k$ series
\begin{subequations} \label{ck}
\begin{equation}
\tilde{c}_{\sigma\sigma'}(k) = -\frac{\beta}{k^2} q_{\sigma} q_{\sigma'}
+ c_{\sigma\sigma'}^{(0)} + c_{\sigma\sigma'}^{(2)} k^2 + O(k^4) , 
\end{equation}
where
\begin{eqnarray}
c_{\sigma\sigma'}^{(0)} & = & \frac{\beta^2}{(2\kappa)^2} q_{\sigma}^2 q_{\sigma'}^2
+ O(\beta^3) , \nonumber \\
c_{\sigma\sigma'}^{(2)} & = & -\frac{2}{3} \frac{\beta^2}{(2\kappa)^4} 
q_{\sigma}^2 q_{\sigma'}^2 + O(\beta^3) .
\end{eqnarray}
\end{subequations}

The correlation function $h_{\sigma\sigma'}(r)\equiv h_{\sigma'\sigma}(r)$ 
is short-ranged due to the screening phenomenon and therefore its small-$k$ 
expansion is regular:
\begin{equation} \label{hk}
\tilde{h}_{\sigma\sigma'}(k) = h_{\sigma\sigma'}^{(0)} + h_{\sigma\sigma'}^{(2)} k^2 
+ h_{\sigma\sigma'}^{(4)} k^4 + O(k^6) .
\end{equation}
According to the 2D Fourier expansion (\ref{Fourier}), the expansion 
coefficients are related to the moments of $h_{\sigma\sigma'}(r)$ as follows
\begin{equation} \label{momentsh}
h_{\sigma\sigma'}^{(2j)} = \frac{(-1)^j}{4^j(j!)^2} 
\int \frac{{\rm d}^2r}{2\pi} r^{2j} h_{\sigma\sigma'}(r) .
\end{equation}
The correlation function is related to the direct correlation function via 
the Ornstein-Zernike (OZ) equation
\begin{equation} \label{OZ} 
\tilde{h}_{\sigma\sigma'}(k) = \tilde{c}_{\sigma\sigma'}(k) + 2\pi \sum_{\sigma''}
\tilde{c}_{\sigma\sigma''}(k) n_{\sigma''} \tilde{h}_{\sigma''\sigma'}(k) .   
\end{equation} 

The next step is to substitute the small-$k$ expansions of the direct
correlation (\ref{ck}) and correlation (\ref{hk}) functions into the
OZ equation (\ref{OZ}).
Setting successively the coefficients to powers of $k$ to 0 implies 
exact constraints for the moments of $h_{\sigma\sigma'}(r)$ and
$c_{\sigma\sigma'}(r)$.
The term of order $1/k^2$ implies the standard zeroth-moment 
screening condition
\begin{equation}
\sum_{\sigma'} q_{\sigma'} n_{\sigma'} h^{(0)}_{\sigma'\sigma} 
= - \frac{q_{\sigma}}{2\pi} .
\end{equation}
The term of order $k^0$ leads to the constraint
\begin{subequations} \label{1}
\begin{equation}
-2\pi\beta\rho_b \sum_{\sigma'} q_{\sigma'} n_{\sigma'} h^{(2)}_{\sigma'\sigma}
= f^{(0)}_{\sigma} + \sum_{\sigma'} \left( 2\pi f^{(0)}_{\sigma'} - 1 \right)
n_{\sigma'} h^{(0)}_{\sigma'\sigma} 
\end{equation}
where
\begin{equation}
f^{(0)}_{\sigma} = \sum_{\sigma'} n_{\sigma'} c^{(0)}_{\sigma'\sigma} 
= \frac{\beta}{8\pi} q_{\sigma}^2 .
\end{equation}
\end{subequations}
The term of order $k^2$ implies
\begin{subequations} \label{2}
\begin{equation} 
-2\pi\beta\rho_b \sum_{\sigma'} q_{\sigma'} n_{\sigma'} h^{(4)}_{\sigma'\sigma}
= f^{(2)}_{\sigma} + \sum_{\sigma'} \left( 2\pi f^{(0)}_{\sigma'} - 1 \right)
n_{\sigma'} h^{(2)}_{\sigma'\sigma} + 2\pi \sum_{\sigma'} f^{(2)}_{\sigma'} 
n_{\sigma'} h^{(0)}_{\sigma'\sigma}
\end{equation}
where
\begin{equation}
f^{(2)}_{\sigma} = \sum_{\sigma'} n_{\sigma'} c^{(2)}_{\sigma'\sigma} 
= - \frac{\beta}{48\pi\kappa^2} q_{\sigma}^2 .
\end{equation}
\end{subequations}

Let us consider a mixture of two species.
Plasma particles correspond to species 1 with charge $q_1=e$ and
number density $n_1=n$ while for dealing with one guest charge $q_2=Z e$
only we set $n_2=0$.
Thus, the background charge density $\rho_b=-n e$ and $\kappa^2 = 2\pi\Gamma n$.
Identifying $\rho({\bf r}\vert Ze,{\bf 0}) = n e h_{21}(r)$, using the
relation (\ref{momentsh}) and setting $\sigma=2$ in the sum rule (\ref{2}), 
one gets
\begin{eqnarray}
\frac{\kappa^2}{4^3 e} \int {\rm d}^2r r^4 \rho({\bf r}\vert Ze,{\bf 0}) 
& = & - \frac{\Gamma}{24\kappa^2} Z^2 - \frac{\Gamma}{24 e \kappa^2}
\int {\rm d}^2r \rho({\bf r}\vert Ze,{\bf 0}) \nonumber \\ & & 
+ \frac{1}{4e} \left( 1 - \frac{\Gamma}{4} \right)
\int {\rm d}^2r r^2 \rho({\bf r}\vert Ze,{\bf 0}) .
\end{eqnarray}
Finally, substituting the previously derived sum rules (\ref{srrr1}) and
(\ref{srrr2}) into this relation we recover the first two terms of 
the small-$\Gamma$ expansion (\ref{res}).

To go to higher-order terms in $\Gamma$ requires very complicated 
calculations as the contributions to the direct correlation function, 
resulting from functional derivatives of completely renormalized diagrams,
are numerous.
As follows from the form of the fourth moment (\ref{fourth}),
the coefficients to powers of $Z$ are finite polynomials in $\Gamma$.
This indicates a cancellation property of certain families of renormalized 
diagrams in the expansion of the direct correlation function, as it was in 
the calculation of the sixth-moment sum rule for the correlation functions 
of the homogeneous 2D one-component plasma \cite{Kalinay00}. 

\renewcommand{\theequation}{4.\arabic{equation}}
\setcounter{equation}{0}

\section{Check at the free-fermion point $\Gamma=2$} \label{Sect4}
For $\Gamma=2$, the induced charge density $\rho(r\vert Ze,{\bf 0})$
was evaluated in Ref. \cite{Samaj07}.
Let us recall the definition of the incomplete Gamma function 
\begin{equation}
\Gamma(s,t) = \int_t^{\infty} {\rm d}x x^{s-1} {\rm e}^{-x}
\end{equation} 
and the Gamma function $\Gamma(s)\equiv \Gamma(s,0)$ \cite{Gradshteyn}.
For positive values of $Z$, one has
\begin{equation}
\rho(r\vert Ze,{\bf 0}) = - e n \frac{\Gamma(Z,\pi nr^2)}{\Gamma(Z)} ,
\qquad Z>0 .
\end{equation} 
For negative values of $Z$, it holds that
\begin{equation}
\rho(r\vert Ze,{\bf 0}) = e n \left[ {\rm e}^{-\pi nr^2} 
\frac{(\pi nr^2)^Z}{\Gamma(Z+1)} - 
\frac{\Gamma(Z+1,\pi nr^2)}{\Gamma(Z+1)} \right] , \qquad -1< Z <0 .
\end{equation} 
In both cases, the formula for even moments of $\rho(r\vert Ze,{\bf 0})$
is the same:
\begin{equation} \label{ff}
\int {\rm d}^2 r r^{2j} \rho(r\vert Ze,{\bf 0}) = - \frac{e}{(j+1) (\pi n)^j}
\frac{\Gamma(Z+j+1)}{\Gamma(Z)} , \qquad j=0,1,2,\ldots.
\end{equation}
In particular,
\begin{equation}
\int {\rm d}^2 r r^4 \rho(r\vert Ze,{\bf 0}) = - \frac{e}{3 (\pi n)^2}
Z (Z+1) (Z+2) .
\end{equation}
This result is consistent with formula (\ref{fourth}) taken at $\Gamma=2$.

\renewcommand{\theequation}{5.\arabic{equation}}
\setcounter{equation}{0}

\section{Sign oscillations of the induced charge density} \label{Sect5}
The knowledge of certain moments can provide an exact information 
about sign oscillations of the induce charge density.
The interaction Boltzmann factor of the guest charge $Z e$ and a plasma
charge $e$ is $r^{\Gamma Z}$.
If $Z>0$, there are no plasma particles at $r=0$ and only the background charge
contributes to the induced charge density, $\rho(0\vert Ze,{\bf 0})=-n e$. 
In the case of a monotonous increase of $\rho(r\vert Ze,{\bf 0})$ from
$-n e$ at $r=0$ to 0 at $r\to\infty$, each integral
$\int {\rm d}^2r r^{2j}\rho(r\vert Ze,{\bf 0}) <0$ $(j=0,1,2,\ldots)$.
On the other hand, the {\em sufficient} condition for sign oscillations
of $\rho(r\vert Ze,{\bf 0})$ is
\begin{equation} \label{Zpositive}
\int {\rm d}^2r r^{2j}\rho(r\vert Ze,{\bf 0}) > 0 , \qquad Z>0 .
\end{equation}
If $Z<0$, the plasma particles are attracted strongly to the guest charge
and $\rho(0\vert Ze,{\bf 0})\to\infty$.
In the case of a monotonous decrease of $\rho(r\vert Ze,{\bf 0})$ from
$\infty$ at $r=0$ to 0 at $r\to\infty$, each integral
$\int {\rm d}^2r r^{2j}\rho(r\vert Ze,{\bf 0}) > 0$ $(j=0,1,2,\ldots)$.
The sufficient condition for the sign oscillations of 
$\rho(r\vert Ze,{\bf 0})$ is
\begin{equation} \label{Znegative}
\int {\rm d}^2r r^{2j}\rho(r\vert Ze,{\bf 0}) < 0 , \qquad Z<0 .
\end{equation}

The zeroth-moment condition (\ref{srrr1}) is not informative,
neither of the conditions (\ref{Zpositive}) and (\ref{Znegative})
with $j=0$ can be satisfied.

The second-moment condition (\ref{srrr2}) was analyzed in Ref. \cite{Samaj07}
and here we only repeat the analysis.
For $Z>0$, the sufficient condition (\ref{Zpositive}) with $j=1$ is
satisfied provided that
\begin{equation} \label{ss}
0 < Z < 1 - \frac{4}{\Gamma} .
\end{equation}
This inequality ensures an interval of positive solutions for $Z$ if 
$\Gamma>4$.
For $Z<0$, the sufficient condition (\ref{Znegative}) with $j=1$ is
identical to the one (\ref{ss}).
Since the collapse phenomenon requires that $Z>-2/\Gamma$, the sufficient
condition for sign oscillations reads as 
\begin{equation} \label{sss}
- \frac{2}{\Gamma} < Z < 1 - \frac{4}{\Gamma} .
\end{equation}
These inequalities have no solution for $Z$ if $\Gamma<2$.
For $\Gamma\ge 4$, $Z$ covers the whole interval of possible negative
values $(-2/\Gamma,0)$.
Combining equations (\ref{ss}) and (\ref{sss}) and taking the sign of $Z$
to be arbitrary, the general condition for $Z$-values with sign 
oscillations of $\rho(r\vert Ze,{\bf 0})$ is the one (\ref{sss}).
Note that the interval (\ref{sss}) does not contain $Z=1$ for any finite
value of $\Gamma$, so the second-moment condition does not provide
any useful information about the sign oscillations of the pair correlation
function of plasma particles themselves. 

We proceed by the analysis of the new fourth-moment condition (\ref{fourth}). 
For $Z>0$, the sufficient condition (\ref{Zpositive}) with $j=2$ is
satisfied provided that
\begin{equation} \label{ZZ}
c_1(\Gamma) + c_2(\Gamma) Z + c_3(\Gamma) Z^2 > 0 .
\end{equation}
This inequality of second degree in $Z$ has the explicit solution
\begin{equation} \label{Zfourth}
Z \in \left( \frac{2\Gamma-7-\vert\Gamma-1\vert}{\Gamma},
\frac{2\Gamma-7+\vert\Gamma-1\vert}{\Gamma} \right) .
\end{equation}
There are two important values of $\Gamma$.
The first one corresponds to the situation when the upper value of $Z$
in (\ref{Zfourth}) intersects 0, $2\Gamma-7+\vert\Gamma-1\vert = 0$,
i.e. $\Gamma=8/3$.
The second one corresponds to the situation when the lower value of $Z$
in (\ref{Zfourth}) intersects 0, $2\Gamma-7-\vert\Gamma-1\vert = 0$,
i.e. $\Gamma=6$.
Thus,
\begin{eqnarray}
Z & \in & \left( 0, 3 - \frac{8}{\Gamma} \right) \qquad
\mbox{for $\Gamma\in\left( \frac{8}{3},6\right)$} \\
Z & \in & \left( 1 - \frac{6}{\Gamma}, 3 - \frac{8}{\Gamma} \right) \qquad
\mbox{for $\Gamma\in\left( 6,\infty\right).$}
\end{eqnarray}
Note that these intervals involve the special value $Z=1$ for $\Gamma>4$, i.e.
the pair correlation function of plasma particles exhibits sign oscillations. 
For $-2/\Gamma < Z< 0$, the sufficient condition (\ref{Znegative}) with $j=2$ 
is identical to the previous one (\ref{ZZ}) with the same solution 
(\ref{Zfourth}).
There is no overlap of this solution with the considered interval 
$-2/\Gamma < Z< 0$ for $\Gamma<2$ and $\Gamma>6$.
The overlap is nonzero only in these three cases:
\begin{eqnarray}
Z & \in & \left( -\frac{2}{\Gamma},3-\frac{8}{\Gamma} \right) \qquad
\mbox{for $2\le \Gamma\le \frac{8}{3}$,} \\
Z & \in & \left( -\frac{2}{\Gamma},0 \right) \qquad
\mbox{for $\frac{8}{3}\le \Gamma\le 4$,} \\
Z & \in & \left( 1 - \frac{6}{\Gamma},0 \right) \qquad
\mbox{for $4\le \Gamma\le 6$.}
\end{eqnarray}

Taking into account all sufficient condition for sign oscillations of
the induced charge cloud, from both second-moment and fourth-moment
sum rules, we get the requirement
\begin{equation}
-\frac{2}{\Gamma} < Z < 3 - \frac{8}{\Gamma} .
\end{equation}
These inequalities have no solution for $\Gamma<2$.
For $2<\Gamma<8/3$, there exist only negative values of $Z$.
For $\Gamma>8/3$, there are also positive values of $Z$ and
they involve the special ``plasma'' value $Z=1$ for $\Gamma>4$. 

The present sufficient conditions for sign oscillations of the charge
density do not exclude the presence of oscillations also for other values 
of $Z$ and $\Gamma$, especially for both $Z$ and $\Gamma$ large.

\renewcommand{\theequation}{6.\arabic{equation}}
\setcounter{equation}{0}

\section{Conclusion} \label{Sect6}
The aim of this paper was to derive the fourth-moment condition
(\ref{fourth}) for the charge density induced around a guest charge
$Z e$ in the two-dimensional jellium.
The derivation was based on the assumption that the fourth moment
is an analytical series in $Z$.
The large-$Z$ analysis performed in Sect. \ref{Sect22} indicates
that the series is finite, up to the $Z^3$ term.
Then the knowledge of the $\Gamma$-dependence of the limit $Z\to 0$ 
(\ref{Zlinear}) and of the special case $Z=1$ (\ref{Zequal}) permits one 
to obtain the fourth moment of the induced charge density (\ref{fourth}).
This result was checked in the Debye-H\"uckel limit $\Gamma\to 0$ and
its first $\Gamma$ correction (Sect. \ref{Sect3}) and at the free-fermion
point $\Gamma=2$ (Sect. \ref{Sect4}).

At $\Gamma=2$, considering in the explicit formula for the $2j$ moment of 
the induced charge density (\ref{ff}) that
\begin{equation}
\frac{\Gamma(Z+j+1)}{\Gamma(Z)} = \prod_{k=0}^j (Z+k) ,
\end{equation}
this moment is the finite polynomial in $Z$ of degree $j+1$.
Since the coupling $\Gamma=2$ is not exceptional from the point of
view of moments it is natural to assume that the $2j$-th moment of the
induced charge density is a finite polynomial in $Z$ of degree $j+1$
for any value of $\Gamma$.
This suggestion is supported by a straightforward extension of the large-$Z$ 
analysis of Sect. \ref{Sect22} to higher moments:
\begin{equation}
\int {\rm d}^2 r r^{2j} \rho({\bf r}\vert Ze,{\bf 0}) \mathop{\sim}_{Z\to\infty}
(-ne) \int_0^R {\rm d}r 2\pi r^{2j+1} = - Z^{j+1} \frac{e}{(j+1) (\pi n)^j} . 
\end{equation} 
This term, which does not depend on $\Gamma$, is easily detectable
in formula (\ref{ff}) for the $2j$-th moment at $\Gamma=2$ due to the relation
\begin{equation}
\frac{\Gamma(Z+j+1)}{\Gamma(Z)} \mathop{\sim}_{Z\to\infty} Z^{j+1} .
\end{equation}
To go beyond the fourth moment is prevented by a complicated 
form of the coefficients of the finite $Z$-series which are probably
no longer finite polynomials in $\Gamma$.
Like for instance, the coefficient to the term linear in $Z$ is not
known for the sixth and higher moments since the expansion of 
$\tilde{S}(k)$ in (\ref{linZ}) is available only up to $k^6$ power or,
in other words, there is no exact result for the 8th moment of the pair
correlation function of plasma particles \cite{Forrester01}.

The derivation of the fourth moment of the charge density around a
guest charge in the 2D jellium was based on plausible, but not rigorously
justified arguments.
Like for instance, the fourth moment (\ref{fourth}) is assumed to be 
a finite polynomial in $Z$, i.e. analytic around $Z=0$, without specifying 
the radius of convergence.
The correct reproduction of the $Z=1$ case and of the limit $Z\to\infty$
indicate that for positive $Z$ the radius of convergence is infinite.
This indication is supported by the fact that the coefficients 
of the polynomial expansion (\ref{fourthb}) are finite for any coupling 
$\Gamma$ of the fluid phase.
As concerns the negative values of $Z$, the collapse phenomenon 
at $Z=-2/\Gamma$ certainly limits the convergence of the finite polynomial
series to $Z>-2/\Gamma$. 
The rigorous validity of the second-moment condition (\ref{srrr2})
for all $Z>-2/\Gamma$ also supports the analogous convergence properties
of the fourth moment.  

An open question is the generalization of the fourth-moment condition
for the 2D jellium to other Coulomb fluids containing various types of
interacting charged species.
In the case of the second-moment, such generalization was made in 2D by
using the BGY hierarchy in Ref. \cite{Jancovici08} and in higher spatial 
dimensions in Ref. \cite{Suttorp08}.
To accomplish this aim one has to go to higher orders of the BGY hierarchy
and to truncate the hierarchy via new symmetries of many-particle correlation 
functions which we were not able to recognize so far.

\begin{acknowledgements}
The support received from Grant VEGA No. 2/0003/18 is acknowledged. 
\end{acknowledgements}


\begin{thebibliography}{10}

\bibitem{Abe59} Abe, R.:
Giant cluster expansion theory and its application to high temperature plasma.
Prog. Theor. Phys. {\bf 22}, 213-226 (1959)

\bibitem{Alastuey81} Alastuey, A., Jancovici, B.:
On the classical two-dimensional one-component Coulomb plasma.
J. Physique {\bf 42}, 1--12 (1981)

\bibitem{Attard02} Attard, Ph.:
Thermodynamics and Statistical Mechanics, Chapter 12. 
Academic Press, London (2002)

\bibitem{Baus78} Baus, M.:
On the compressibility of a one-component plasma.
J. Phys. A: Math. Gen. {\bf 11}, 2451--2462 (1978)

\bibitem{Baus80} Baus, M., Hansen J.P.:
Statistical mechanics of simple Coulomb systems.
Phys. Rep. {\bf 59}, 1--94 (1980)

\bibitem{DiFrancesco94} Di Francesco, P., Gaudin, M., Itzykson, C., Lesage, F.:
Laughlin's wave functions, Coulomb gases and expansions of the discriminant.
Int. J. Mod. Phys. A {\bf 9}, 4257--4351 (1994)

\bibitem{Deutsch74} Deutsch, C., Lavaud, M.:
Equilibrium properties of a two-dimensional Coulomb gas.
Phys. Rev. A {\bf 9}, 2598--2616 (1974)

\bibitem{Forrester98} Forrester, P.J.:
Exact results for two-dimensional Coulomb systems.
Phys. Rep. {\bf 301}, 235--270 (1998)

\bibitem{Forrester01} Forrester, P.J., Jancovici, B., McAnally, D.S.:
Analytic properties of the structure function for the one-dimensional
one-component log-gas.
J. Stat. Phys. {\bf 102}, 737--780 (2001)

\bibitem{Friedman59} Friedman, H.L.:
On Mayer's ionic solution theory.
Mol. Phys. {\bf 2}, 23-38 (1959)

\bibitem{Gradshteyn} Gradshteyn, I.S., Ryzhik, I.M.:
Table of Integrals, Series and Products, 5th. edn.
Academic Press, London (1994)

\bibitem{Jancovici81a} Jancovici, B.:
Exact results for the two-dimensional one-component plasma.
Phys. Rev. Lett. {\bf 46}, 386--388 (1981)

\bibitem{Jancovici81b} Jancovici, B.:
Charge distribution and kinetic pressure in a plasma: a soluble model.
J. Physique Lett. {\bf 42}, 223--226 (1981)

\bibitem{Jancovici82} Jancovici, B.:
Classical Coulomb systems near a plane wall. I.
J. Stat. Phys. {\bf 28}, 43--65 (1982)

\bibitem{Jancovici92} Jancovici, B.:
Inhomogeneous two-dimensional plasmas.
In: Henderson. D. (ed.) Inhomogeneous Fluids, pp. 201--237, Dekker, 
New York (1992)

\bibitem{Jancovici00b} Jancovici, B., Kalinay, P., \v{S}amaj, L.:
Another derivation of a sum rule for the two-dimensional two-component plasma.
Physica A {\bf 279}, 260--267 (2000)

\bibitem{Jancovici08} Jancovici, B., \v{S}amaj, L.:
Guest charge and potential fluctuations in two-dimensional
classical Coulomb systems.
J. Stat. Phys. {\bf 131}, 613--629 (2008)

\bibitem{Kalinay00} Kalinay, P., Marko\v{s}, P., \v{S}amaj, L., 
Trav\v{e}nec, I.:
The sixth-moment sum rule for the pair correlations of the two-dimensional
one-component plasma: Exact result.
J. Stat. Phys. {\bf 98}, 639--666 (2000)

\bibitem{Martin88} Martin, Ph.A.:
Sum rules in charged fluids.
Rev. Mod. Phys. {\bf 60}, 1075--1127 (1988)

\bibitem{Meeron61} Meeron, E.:
Plasma Physics. Mac Graw-Hill, New York (1961)

\bibitem{Mora15} Mora Grimaldo, J.A.M., T\'ellez, G.: 
Relations among two methods for computing the partition function
of the two-dimensional one-component plasma.
J. Stat. Phys. {\bf 160}, 4--28 (2015)

\bibitem{Salazar16} Salazar, R., T\'ellez, G.: 
Exact energy computation of the one component plasma on a sphere for
even values of the coupling parameter,
J. Stat. Phys. {\bf 164}, 969--999 (2016)

\bibitem{Samaj95} \v{S}amaj, L., Percus, J.K.:
A functional relation among the pair correlations of the two-dimensional
one-component plasma.
J. Stat. Phys. {\bf 80}, 811--824 (1995)

\bibitem{Samaj00} \v{S}amaj, L., Trav\v{e}nec, I.:
Thermodynamic properties of the two-dimensional two-component plasma.
J. Stat. Phys. {\bf 101}, 713--730 (2000)

\bibitem{Samaj04} \v{S}amaj, L.:
Is the two-dimensional one-component plasma exactly solvable?
J. Stat. Phys. {\bf 117}, 131--158 (2004)

\bibitem{Samaj07} \v{S}amaj, L.:
A generalization of the Stillinger-Lovett sum rules for the two-dimensional
jellium.
J. Stat. Phys. {\bf 128}, 1415--1428 (2007)

\bibitem{Stillinger68a} Stillinger, F.H., Lovett, R.:
Ion-pair theory of concentrated electrolytes. I. Basic Concepts.
J. Chem. Phys. {\bf 48}, 3858 (1968)

\bibitem{Stillinger68b} Stillinger, F.H., Lovett, R.:
General restriction on the distribution of ions in electrolytes.
J. Chem. Phys. {\bf 49}, 1991 (1968)

\bibitem{Suttorp08} Suttorp, L.G.:
Sum rules for correlation functions of ionic mixtures in arbitrary
dimension $d\ge 2$.
J. Phys. A: Math. Theor. {\bf 41}, 495001 (2008)

\bibitem{Tellez99} T\'ellez, G., Forrester, P.J.:
Exact finite-size study of the 2D OCP at $\Gamma=4$ and $\Gamma=6$.
J. Stat. Phys. {\bf 97}, 489--521 (1999)

\bibitem{Tellez12} T\'ellez, G., Forrester, P.J.:
Expanded Vandermonde powers and sum rules for the two-dimensional
one-component plasma.
J. Stat. Phys. {\bf 148}, 824--855 (2012)

\bibitem{Vieillefosse75} Vieillefosse, P., Hansen, J.P.:
Statistical mechanics of dense ionized matter. V. Hydrodynamic limit and 
transport coefficients of the classical one-component plasma.
Phys. Rev. A {\bf 12}, 1106--1116 (1975)

\bibitem{Vieillefosse85} Vieillefosse, P.:     
Sum rules and perfect screening conditions for the one-component plasma
J. Stat. Phys. {\bf 41}, 1015--1035 (1985)

\end{thebibliography}
\end{document}